\documentclass[superscriptaddress,secnumarabic,
amssymb,amsmath,nobibnotes,aps,prd,showkeys,showpacs,nofootinbib]{revtex4}%
\usepackage{graphicx}
\usepackage{epsf}
\usepackage{bm}
\usepackage{amsmath}
\usepackage{amsfonts}
\usepackage{amssymb}
\usepackage{epstopdf}
\usepackage{natbib}%
\providecommand{\U}[1]{\protect\rule{.1in}{.1in}}
\newcommand{\be}{\begin{equation}}
\newcommand{\ee}{\end{equation}}

\newcommand{\mincir}{\raise
-3.truept\hbox{\rlap{\hbox{$\sim$}}\raise4.truept\hbox{$<$}\ }}
\newcommand{\magcir}{\raise
-3.truept\hbox{\rlap{\hbox{$\sim$}}\raise4.truept\hbox{$>$}\ }}

\begin{document}
\title{An analytic model for interacting dark energy and its observational constraints}
\author{Supriya Pan}
\email{span@research.jdvu.ac.in}
\affiliation{Department of Mathematics, Jadavpur University, Kolkata-700032, India}
\author{Subhra Bhattacharya}
\email{subhra.maths@presiuniv.ac.in}
\affiliation{Department of Mathematics, Presidency University, Kolkata-700073, India}
\author{Subenoy Chakraborty}
\email{schakraborty@math.jdvu.ac.in}
\affiliation{Department of Mathematics, Jadavpur University, Kolkata-700032, India}
%%%%%%%%%%%%%%%%%%%%%%%%%%%%%%%%%%%%%%%%%%%%%%%%%%%%%%%%%%%%%%%%%%%%%%%%%%%%%%%%%%%%%%%%%%%%%%%%%%%%%%%%%%%%%%%%%%
%%%%%%%%%%%%%%%%%%%%%%%%%%%%%%%%%%%%%%%%%%%%%%%%%%%%%%%%%%%%%%%%%%%%%%%%%%%%%%%%%%%%%%%%%%%%%%%%%%%%%%%%%%%%%%%%%%
\begin{abstract}

The paper deals with a theoretical model for interacting dark energy. The interaction between
the cold dark matter (dust) and the dark energy has been assumed to be non-gravitational in
nature. Exact analytic cosmological solutions are obtained both for constant and variable
equation of state for dark energy. It is found that, for very small value of the coupling
parameter (in the interaction term), the model asymptotically extends up to $\Lambda$CDM,
while the model can enter into the phantom domain asymptotically, if the coupling parameter
is not so small. Both the solutions are then analyzed with 194 Supernovae Type Ia data. The best fit
parameters are shown with 1$\sigma$ and 2$\sigma$ confidence intervals. Finally, we have
discussed the cosmographic parameters for both the cases.
\end{abstract}

%%%%%%%%%%%%%%%%%%%%%%%%%%%%%%%%%%%%%%%%%%%%%%%%%%%%%%%%%%%%%%%%%%%%%%%%%%%%%%%%%%%%
\maketitle
%%%%%%%%%%%%%%%%%%%%%%%%%%%%%%%%%%%%%%%%%%%%%%%%%%%%%%%%%%%%%%%%%%%%%%%%%%%%%%%%%%%%%%%%%%%%%%%%%%%%%%%%%%%%%%%%%%
%~~~\myclassification{98.80.Cq, 98.80.-k}\\\\
%%%%%%%%%%%%%%%%%%%%%%%%%%%%%%%%%%%%%%%%%%%%%%%%%%%%%%%%%%%%%%%%%%%%%%%%%%%%%%%%%%%%%%%%%%%%%%%%%%%%%%%%%%%%%%%%%%
\section{Introduction}

There are lot of observational evidences which prove that our universe is accelerating
at present \cite{Riess1, Perlmutter1, Spergel1, Tegmark1, Eisenstein1, Jain1}.
To explain this acceleration within the framework of Einstein's general relativity,
usually dark energy (DE) having negative pressure is introduced. To search for dark
energy candidates, cosmological constant $\Lambda$ was proposed to be the simplest
candidate for dark energy, and, combining it with cold dark matter (CDM), the model
$\Lambda$CDM sounds good with most of the observational data. But, it became
an embarrassing issue because, it suffers from the cosmological constant or
fine tuning problem \cite{Weinberg1, Carroll1}, and the cosmic coincidence
problem \cite{Copeland1}. Further, recent observations predict that nearly
$73\%$ of our universe is filled with dark energy and $23 \%$ by dark matter
(DM), and the rest 4\% is the usual baryonic matter and radiation. So, people
are trying to find some suitable dark energy candidate which
are free from the above two problems.
As a result, various dark energy models were introduced, such as, Quintessence
\cite{Caldwell1}, K-essence \cite{Armendariz-Picon1}, Tachyon
\cite{Padmanabhan1}, Phantom \cite{Caldwell2}, Quintom
\cite{Elizalde1, Feng1, Cai1}, Chaplygin gas \cite{Kamenshchik1},
Holographic DE \cite{Cohen1, Li1}, and, so on. Still, the nature of
dark energy is still elusive.

Now, to alleviate the cosmic coincidence problem, and, to know the nature of DE,
people introduced interacting dynamics between DM and DE \cite{Wetterich-ide1,
Amendola-ide1, Amendola-ide2, Billyard-ide1, Zimdahl-ide1, Herrera-ide1,
Amendola-ide3, Hoffman-ide1, Chimento-ide1, Amendola-ide4, AP2014}. Although,
there is no strong reason to exclude this dynamics, but, still the
question arises, what should be the possible form of the interaction
between these two components. As there is no such well motivated arguments
behind this choice, so, we deal with phenomenological interacting term between
these components, but sometimes, we choose from mathematical point of view, or,
we constrain the parameters in the interaction terms by latest data.

In this work, we have considered dark energy interacting with cold dark matter
by some phenomenological interaction term between them.
The cases for constant and variable dark energy equation of state ($\omega_d$)
have been investigated. It is interesting to mention that,
the case for variable EoS with very very small interaction between DM and DE
leads to the $\Lambda$CDM ($\omega_d= -1$) model at late-time, but for large interaction (though ``$< 1$'')
between these dark sectors, EoS can cross the phantom divide line, i.e, $\omega_d< -1$.
This is the familiar characteristic of the quintom models \cite{Elizalde1, Feng1,
Cai1}, and some other model in different context of cosmology \cite{Cai2007, Pan2014}
already existing in the literature. The restriction in the variable EoS thus
contain some noteworthy properties in the cosmic history. We have analyzed our
both the models with 194 Supernovae data by Tonry et al. \cite{Tonry2003}
and Barris et al. \cite{Barris2004}. Finally, we have shown the variation of the
cosmographic parameters graphically throughout the entire evolution
of the universe in the context of interacting dark energy model.

The paper has been organized in this way: In Section 2, we have presented the interaction dynamics
between the dark sectors: dark matter and dark energy. We have tried to find analytic solutions
both for constant and variable EoS for dark energy. In Section 3, we have analyzed our
model by 194 Supernovae data. Section 4 contains the cosmographic analysis for both the models.
Finally, we have presented a brief summary in Section 5.

\section{Interacting Dark sectors: Tracing the Cosmic history}

Consider that our universe is well described by a flat Friedmann--Lemaitre--Robertson--Walker (FLRW) line element

\begin{equation}
ds^2= -dt^2+ a^2 (t) (dx^2+ dy^2+ dz^2),\label{FLRW-metric}
\end{equation}
and, the matter distribution obeys the perfect fluid distribution with the energy-momentum tensor

\begin{equation}
T_{\mu \nu}= (p+ \rho) u_{\mu} u_{\nu}+ p g_{\mu \nu},\label{energy-momentum}
\end{equation}
where $u_{\mu}$ is the four velocity vector of the perfect fluid, $\rho$, $p$ are
the energy density and the thermodynamic pressure of the perfect fluid. Thus, the explicit
form of the Einstein's field equations (assuming $c= 1$)

\begin{equation}
G_{\mu \nu}= 8 \pi G T_{\mu \nu},\label{Einstein-equation}
\end{equation}
are the Friedmann's equations
%%%%%%%%%%%%%%%%%%%%%%%%%%%%%%%%%%%%%%%%%%%%%%%%%%%%%%%%%%%%%%%%%%%%
\begin{eqnarray}
H^2&=&\frac{8\pi G}{3}(\rho_m+\rho_d),\label{friedmann1}\\
2\dot{H}+ 3 H^2&=& - 8 \pi G (p_m+ p_d),\label{friedmann2}
\end{eqnarray}
%%%%%%%%%%%%%%%%%%%%%%%%%%%%%%%%%%%%%%%%%%%%%%%%%%%%%%%%%%%%%%%%%%
where $H= \dot{a}/ a$, is the Hubble parameter, an overdot
represents the differentiation with respect to the cosmic time $t$, $\rho_m$, $\rho_d$
are the energy densities of DM and DE, and $p_m$, $p_d$ are the corresponding thermodynamic pressures of the
two dark components. Further, we assume that the dark matter is in the form of a pressureless dust (i.e., $p_m= 0$)
and the dark energy satisfies the barotropic equation of state $p_d= \omega_d \rho_d$, where $\omega_d$ is
the equation of state for dark energy. Thus considering the interaction between these two
components, we can write the conservation equations both for DM and DE
in the following coupled form:

\begin{eqnarray}
\dot{\rho_m}+3 H \rho_m&=& Q,\label{conservation1}\\
\dot{\rho_d}+3 H (1+\omega_d) \rho_d&=& -Q.\label{conservation2}
\end{eqnarray}
%%%%%%%%%%%%%%%%%%%%%%%%%%%%%%%%%%%%%%%%%%%%%%%%%%%%%%%%%%%%%%%%%%%%%%%%%%%%%%%%%%%%%
Here, $Q$ is the rate of energy density exchange between DM and DE, where\\

$\bullet$ $Q> 0$ $\Longrightarrow$ Energy goes from DE to DM,

$\bullet$ $Q< 0$ $\Longrightarrow$ Energy goes from DM to DE.\\

%%%%%%%%%%%%%%%%%%%%%%%%%%%%%%%%%%%%%%%%%%%%%%%%%%%%%%%%%%%%%%%%%%%%%%%%%%

We shall assume $Q$ to be positive for the validity of the second law of
thermodynamics. If we see the continuity Eqns. (\ref{conservation1}) and (\ref{conservation2}),
the interaction between DE and DM must be a function of the energy densities
multiplied by a quantity having units of the inverse of time which has the
natural choice as the Hubble parameter. Thus interaction between DE and DM
could be expressed phenomenologically in the forms, such as, (i) $Q= Q(H \rho_m)$,
(ii) $Q= Q(H \rho_d)$, (iii) $Q= Q[H (\rho_d+ \rho_m)]$, or, more generally,
(iv) $Q= Q(H \rho_d, H \rho_m)$. In the literature, the nature of DE has been
studied considering different type of interactions (for details, see Ref. \cite{He1}).
We consider for simplicity that the interaction is in linear combinations
of the dark sector densities as \cite{Quartin1}

\begin{equation}
Q= 3\lambda_m H \rho_m +3\lambda_d H \rho_d,\label{interaction1}
\end{equation}
%%%%%%%%%%%%%%%%%%%%%%%%%%%%%%%%%%%%%%%%%%%%%%%%%%%%%%%%%%%%%%%%%%%%%%%%%%%
where $\lambda_m$ and $\lambda_d$ are dimensionless constants. As from observational
point of view the interaction should be subdominant today \cite{Chimento1},
so, $|\lambda_m|$ and $|\lambda_d|$ are very small (i.e., $|\lambda_m|$ $\ll$ 1
and $|\lambda_d|$ $\ll$ 1). The factor `3' in the above expression for interaction
is motivated purely from mathematical ground. This general form of interaction has
been studied recently by several authors \cite{Quartin1} and the particular
cases $\lambda_m= \lambda_d$, in Ref. \cite{Chimento-ide1}, and $\lambda_d= 0$,
in Ref. \cite{Billyard-ide1}. Inserting Eq. (\ref{interaction1}) in the energy 
conservation Eqns. (\ref{conservation1}) and (\ref{conservation2}) we have

\begin{eqnarray}
\dot{\rho_m}+ 3 H \left(1-\lambda_m-\frac{\lambda_d}{u}\right)\rho_m &=&0,\label{conservation1.1}\\
\dot{\rho_d}+ 3 H \left(1+\omega_d+\lambda_d+\lambda_m u\right)\rho_d&=&0, \label{conservation2.1}
\end{eqnarray}
%%%%%%%%%%%%%%%%%%%%%%%%%%%%%%%%%%%%%%%%%%%%%%%%%%%%%%%%%%%%%%%%%%%%%%%%%%%%%%%%%%%%%%%%%%%%%%%%%%%%%%
where $u=\rho_m/\rho_d$. Eqns. (\ref{conservation1.1}) and (\ref{conservation2.1})
show that we have effectively non-interacting two fluid system, where both the
components have the energy densities as before, only pressure changes. If we
define $\rho_t =\rho_m+\rho_d$, as the total energy density of the combined fluid,
then its evolution equation can be obtained from the conservation relations
(either Eqns. (\ref{conservation1}) and (\ref{conservation2}), or Eqns.
(\ref{conservation1.1}) and (\ref{conservation2.1})) as

\begin{equation}
\dot{\rho_t}= -3 H \rho_m- 3 H (1+\omega_d)\rho_d~~~~\Longrightarrow~~~~\dot{\rho_t}+ 3 H (1+\omega_t)\rho_t= 0, \label{total-conservation}
\end{equation}
%%%%%%%%%%%%%%%%%%%%%%%%%%%%%%%%%%%%%%%%%%%%%%%%%%%%%%%%%%%%%%%%%%%%%%%%%%%%%%%%%%%%%%%%%%%%%%%%5
with the effective equation of state ($\omega_t$) of the combined fluid as

\begin{equation}
\omega_t = \frac{\omega_d \rho_d}{\rho_t}= \omega_d \Omega_d,\label{combined-eos}
\end{equation}
%%%%%%%%%%%%%%%%%%%%%%%%%%%%%%%%%%%%%%%%%%%%%%%%%%%%%%%%%%%%%%%%%%%%%%%%%%%%%%%%%%%%%%%%%%%%%%%%%%
where $\Omega_d= \rho_d/\rho_c$, ($\rho_c= 3 H^2/ 8 \pi G$, the critical energy density)
is the density parameter of the dark energy which is related to the density parameter
for dark matter ($\Omega_m= \rho_m/ \rho_c$) by the following relation
(a different look of the Eq. (\ref{friedmann1}))

\begin{eqnarray}
\Omega_t &\equiv& \Omega_d+\Omega_m =1.\label{friedmann1.1}
\end{eqnarray}
%%%%%%%%%%%%%%%%%%%%%%%%%%%%%%%%%%%%%%%%%%%%%%%%%%%%%%%%%%
It should be noted that, if $\omega_d$ is chosen to be a constant, $\omega_t$
still be a variable, i.e., the effective one fluid model has always varying
equation of state. According to present observations, ``$\omega_d< -1$''
\cite{Komatsu1, Planck-collaboration}. So, from Eq. (\ref{combined-eos}),
we see that $\omega_t< - \Omega_d$, which shows that for $1/3<\Omega_d<1$,
the equation of state for combined fluid describes a dark energy Universe.
Now, using Eq. (\ref{total-conservation}), we can
solve for $\rho_d$ and $\rho_m$ in the following way:

\begin{eqnarray}
\rho_d =-\left(\frac{\rho_t+ \rho^\prime _t}{\omega_d}\right),\label{DE-density}\\
\rho_m=\left(\frac{\rho^\prime _t+(1+ \omega_d)\rho_t}{\omega_d}\right),\label{DM-density}
\end{eqnarray}
%%%%%%%%%%%%%%%%%%%%%%%%%%%%%%%%%%%%%%%%%%%%%%%%%%%%%%%%%%%%%%%%%%%%%%%%%%%%%%%%%%%%%%%%%%%%%%%%%%
where $^\prime$ represents the differentiation with respect to $x= 3~ln~a$. 
Now, eliminating $\rho_d$ from Eqns. (\ref{conservation2.1})
and (\ref{DE-density}), we obtain a second order differential equation for $\rho_t$ as

\begin{equation}
\rho^{\prime\prime} _t+ \left(2+\omega_d+\lambda_d-\lambda_m-\frac{\omega^\prime _d}{\omega_d}\right)\rho^\prime _t+ \left[(1+\omega_d)(1-\lambda_m)+\lambda_d-\frac{\omega^\prime _d}{\omega _d}\right]\rho_t= 0. \label{diffeqn}
\end{equation}

We shall solve $\rho_t$ for both constant and variable $\omega_d$.\\

$\bullet$ {\bf When $\omega_d$ is assumed to be a constant}\\

For constant equation of state for dark energy, the explicit form of $\rho_t$ is given by

\begin{equation}
\rho_t= \rho_0 (1+ z)^{-3 \mu_0}+ \rho_1 (1+ z)^{3 \mu_1},\label{total}
\end{equation}
%%%%%%%%%%%%%%%%%%%%%%%%%%%%%%%%%%%%%%%%%%%%%%%%%%%%%%%%%%%%%%%%%%%%%%%%%%%%%%%%%%%%%%%%%%%
where $\rho_0$, $\rho_1$ are constants of integration, and,

$$\mu_0= \frac{1}{2} \left[-(2+\omega_d+\lambda_d-\lambda_m)+\sqrt{(\lambda_m+\omega_d+\lambda_d)^2-4\lambda_m \lambda_d}\right],$$

$$\mu_1= \frac{1}{2} \left[2+\omega_d+\lambda_d-\lambda_m+\sqrt{(\lambda_m+\omega_d+\lambda_d)^2-4\lambda_m \lambda_d}\right].$$
%%%%%%%%%%%%%%%%%%%%%%%%%%%%%%%%%%%%%%%%%%%%%%%%%%%%%%%%%%%%%%%%%%%%%%%%%%%%%%%%%%%%%%%%%%%%%%%%%%
The solution for $\rho_t$ in Eq. (\ref{total}) is contained in the works by Chimento \cite{Chimento1} in the 
context of interacting dark energy. As  $|\lambda_m| \ll 1$ and $|\lambda_d| \ll 1$, so we neglect the product term
`$\lambda_m \lambda_d$' within the square root compared to the first term, then
$\mu_0$ and $\mu_1$ are simplified to $\mu_0 \simeq -(1-\lambda_m)$, and,
$\mu_1 \simeq (1+ \omega_d+ \lambda_d)$. Thus using the above approximations
on $\mu_0$ and $\mu_1$ in (\ref{total}), we have

\begin{equation}
\rho_t=3H^2= \rho_0 (1+ z)^{3 (1-\lambda_m)}+ \rho_1 (1+z)^{3 (1+ \omega_d+ \lambda_d)}.\label{total-approx}
\end{equation}
%%%%%%%%%%%%%%%%%%%%%%%%%%%%%%%%%%%%%%%%%%%%%%%%%%%%%%%%%%%%%%%%%%%%%%%%%
The above equation (\ref{total-approx}) shows that, the present interacting
dark matter and dark energy model is equivalent to a non-interacting two fluid model with
constant equation of state parameters `$- \lambda_m$' and `$(\lambda_d+ \omega_d)$'
Also, the integration constants $\rho_0$ and $\rho_1$ can be interpreted as the 
present energy densities of the two equivalent fluid components.
Now, using (\ref{DE-density}) and (\ref{DM-density}), the energy densities of the
two dark species respectively take the form

\begin{eqnarray}
\rho_d&=& \rho_1 \left(\frac{\lambda_d+ \omega_d}{\omega_d}\right) (1+z)^{3(1+\omega_d+\lambda_d)}-\rho_0 \left(\frac{\lambda_m}{\omega_d}\right)(1+z)^{3(1-\lambda_m)},\label{DE-density1}\\
\rho_m&=& \rho_0 \left(\frac{\lambda_m+ \omega_d}{\omega_d}\right) (1+z)^{3(1-\lambda_m)}-\rho_1 \left(\frac{\lambda_d}{\omega_d}\right) (1+z)^{3(1+\omega_d+\lambda_d)},\label{DM-density1}
\end{eqnarray}
%%%%%%%%%%%%%%%%%%%%%%%%%%%%%%%%%%%%%%%%%%%%%%%%%%%%%%%%%%%%
It should be noted that, the solutions for DE [Eq. (\ref{DE-density1})] and DM [Eq. (\ref{DM-density1})]
were exactly found by Chimento \cite{Chimento1}. In connection with the analytic solutions, one may 
notice the analytic solutions for scalar field models in \cite{Lima2008, AP2015}.
Now, using the Eqns. (\ref{DE-density1}) and (\ref{DM-density1}), we can find the present (i.e., at $z= 0$)
energy densities for DE ($\rho_{d0}$) and DM ($\rho_{m0}$). Further, 
we introduce the present values of the density parameters for dark energy ($\Omega_{d0}$) and 
dark matter ($\Omega_{m0}$) respectively as

\begin{eqnarray}
\Omega_{d0}&=& \frac{\rho_{d0}}{3 H_0 ^2}= \Omega_1 \left(\frac{\lambda_d+ \omega_d}{\omega_d}\right)- \Omega_0 \left(\frac{\lambda_m}{\omega_d}\right),\label{neweqn1}\\
\Omega_{m0}&=& \frac{\rho_{m0}}{3 H_0 ^2}= \Omega_0 \left(\frac{\lambda_m+ \omega_d}{\omega_d}\right)-\Omega_1 \left(\frac{\lambda_d}{\omega_d}\right),\label{neweqn2}
\end{eqnarray}
%%%%%%%%%%%%%%%%%%%%%%%%%%%%%%%%%%%%%%%%%%%%%%%%%%%%%%%%%%%%%%%%%%%%%%%%%%%%%%%%%%%%%%%%
where $\Omega_0= \rho_0/ 3 H_0 ^2$, $\Omega_1= \rho_1/ 3 H_0 ^2$, and, also, we see that, $\Omega_{m0}+ \Omega_{d0}= \Omega_0+ \Omega_1= 1$. Specifically, the quantities, $\Omega_0$ and $\Omega_1$ can be expressed as

\begin{eqnarray}
\Omega_0&=& \frac{\lambda_d+ \Omega_{m0} \omega_d}{\lambda_m+ \lambda_d+ \omega_d},\label{neweqn3}\\
\Omega_1&=& \frac{\lambda_m+ \Omega_{d0} \omega_d}{\lambda_m+ \lambda_d+ \omega_d},\label{neweqn4}
\end{eqnarray}
%%%%%%%%%%%%%%%%%%%%%%%%%%%%%%%%%%%%%%%%%%%%%%%%%%%%%%%%%%%%%%%%%%%%%%%%%%%%%%%
Furthermore, from equation (\ref{combined-eos}) we have

\begin{equation}
\omega_t= \frac{(\lambda_d+ \omega_d) \left(1+ z\right)^{3 (\omega_d+ \lambda_m+ \lambda_d)}- \lambda_m\frac{\rho_0}{\rho_1}}{\left(1+z\right)^{3(\omega_d+\lambda_m+\lambda_d)}+ \frac{\rho_0}{\rho_1}}.\label{combined-eos1}
\end{equation}
%%%%%%%%%%%%%%%%%%%%%%%%%%%%%%%%%%%%%%%%%%%%%%%%%%%%%%%%%%%%%%%%%%%%%%%%%%%%%%%%%%
As the energy densities for both the dark components will be positive
throughout the evolution, so, from Eqns. (\ref{DE-density1}) and (\ref{DM-density1}),
we must have, $0 < \mbox{max}(\lambda_m, \lambda_d) < |\omega_d|$. From equation (\ref{combined-eos1}),
we can identify the behavior of the combined equation of state of the dark sector as follows:\\

{\bf I:} $\omega_d+ \lambda_m+ \lambda_d> 0$

\begin{eqnarray}
\mbox{As}~z &\longrightarrow& \infty,~\omega_t \longrightarrow (\omega_d+ \lambda_d),\label{combined-eos1.1}\\
\mbox{As}~z&\longrightarrow& 0,~\omega_t \longrightarrow \frac{\lambda_d+ \omega_d-\rho_0 \lambda_m/\rho_1}{1+ \rho_0/\rho_1},\label{combined-eos1.2}\\
\mbox{As}~z &\longrightarrow&-1, \omega_t ~\mbox{is undefined}.\label{combined-eos1.3}
\end{eqnarray}

{\bf II:} $\omega_d+ \lambda_m+ \lambda_d< 0$

\begin{eqnarray}
\mbox{As}~z &\longrightarrow& \infty,~\omega_t ~\mbox{is undefined},\label{combined-eos1.4}\\
\mbox{As}~z&\longrightarrow& 0,~\omega_t \longrightarrow \frac{\lambda_d+ \omega_d-\rho_0 \lambda_m/\rho_1}{1+ \rho_0/\rho_1},\label{combined-eos1.5}\\
\mbox{As}~z &\longrightarrow&-1, \omega_t \longrightarrow (\omega_d+ \lambda_d).\label{combined-eos1.6}
\end{eqnarray}
%%%%%%%%%%%%%%%%%%%%%%%%%%%%%%%%%%%%%%%%%%%%%%%%%%%%%%%%%%%%%%%%%%%%%%%%
The deceleration parameter $q$ is given by \cite{Pan2013}

\begin{eqnarray}
q&=&\frac{1}{2}+ \frac{3}{2}\Omega_d \omega_d=\frac{1}{2}(1+ 3 \omega_t)= \frac{1}{2} \left [\frac{(1- 3\lambda_m)\frac{\rho_0}{\rho_1}+(1+ 3\lambda_d+ 3\omega_d)(1+ z)^{3(\omega_d+ \lambda_m+ \lambda_d)}}{\frac{\rho_0}{\rho_1}+(1+ z)^{3(\omega_d+ \lambda_m+ \lambda_d)}}\right],\label{deceleration-constant}
\end{eqnarray}
%%%%%%%%%%%%%%%%%%%%%%%%%%%%%%%%%%%%%%%%%%%%%%%%%%%%%%%%%%%%%%%%%%%%%%%%%%%%%%%
and, consequently, in both cases I and II, the deceleration parameter
$q_I$ (for case I) and $q_{II}$ (for case II) in the limit can be viewed as follows:

\begin{eqnarray}
\mbox{As}~~z &\longrightarrow& \infty:~~q_I \longrightarrow \frac{1}{2}\left(1+ 3\omega_d+ 3\lambda_d\right),~~\mbox{and,}~~q_{II}~~\mbox{is undefined},\label{dp-new1}\\
\mbox{As}~~z &\longrightarrow& 0:~~q_I= q_{II}= \frac{1}{2}\left[1+ 3 \left(\frac{\lambda_d+ \omega_d-\rho_0 \frac{\lambda_m}{\rho_1}}{1+ \frac{\rho_0}{\rho_1}}\right)\right],\label{dp-new2}\\
\mbox{As}~~z &\longrightarrow& -1:~~q_I~~\mbox{is undefined},~~\mbox{but,}~~q_{II}=\frac{1}{2} \left(1+ 3\omega_d+ 3\lambda_d\right).\label{dp-new3}
\end{eqnarray}

$\bullet$ {\bf When $\omega_d$ is a variable}\\

The differential equation (\ref{diffeqn}) for $\rho_t$ can not be solved for arbitrary
variation of the equation of state ($\omega_d$) for dark energy. We consider the case
when $\lambda_m= 0$. Thus the Eq. (\ref{diffeqn}) becomes

\begin{equation}
\rho_t ^{\prime \prime}+ \left(2+ \omega_d+ \lambda_d- \frac{\omega_d ^\prime}{\omega_d}\right) \rho_t ^\prime+ \left(1+ \omega_d+ \lambda_d- \frac{\omega_d ^\prime}{\omega_d}\right) \rho_t= 0, \label{variable1}
\end{equation}

Further, we assume that the variation of $\omega_d$ to be restricted by the relation

\begin{equation}
\frac{\omega_d^\prime}{\omega_d}= n+ \omega_d+ \lambda_d, \label{variable2}
\end{equation}
%%%%%%%%%%%%%%%%%%%%%%%%%%%%%%%%%%%%%%%%%%%%%%%%%%%%%%%%%%%%%%%%%%%%%%%%%%%%%%%%%%
where $n$ is any real number. Thus, the solution for $\rho_t$ becomes

\begin{equation}
\rho_t= \rho^0 \left(1+ z\right)^3+ \rho^1 \left(1+ z\right)^{3 (1-n)},\label{variable3}
\end{equation}
%%%%%%%%%%%%%%%%%%%%%%%%%%%%%%%%%%%%%%%%%%%%%%%%%%%%%%%%%%%%%%%%%%%%%%%%%%%%%%%%%%%
where $\rho^0$ and $\rho^1$ are integration constants. Now, depending on the
values of $n$, $\rho_t$ behaves in the following way:\\

$\star$ {\bf $n> 1$:}\\

~~~~~~~~~~~~~~~~~~~~~~~~~~~~~~$\rho_t \longrightarrow \infty,~~\mbox{as},~~z \longrightarrow \infty$,\\

~~~~~~~~~~~~~~~~~~~~~~~~~~~~~~$\rho_t \longrightarrow \rho^0+ \rho^1,~~\mbox{as},~~z \longrightarrow 0$,\\

~~~~~~~~~~~~~~~~~~~~~~~~~~~~~~$\rho_t \longrightarrow \infty,~~\mbox{as},~~z \longrightarrow -1$.\\

$\star$ {\bf $n< 1$:}\\

~~~~~~~~~~~~~~~~~~~~~~~~~~~~~~$\rho_t \longrightarrow \infty,~~\mbox{as},~~z \longrightarrow \infty$,\\

~~~~~~~~~~~~~~~~~~~~~~~~~~~~~~$\rho_t \longrightarrow \rho^0+ \rho^1,~~\mbox{as},~~z \longrightarrow 0$,\\

~~~~~~~~~~~~~~~~~~~~~~~~~~~~~~$\rho_t \longrightarrow 0,~~\mbox{as},~~z \longrightarrow -1$.\\

$\star$ {\bf $n= 1$:}\\

~~~~~~~~~~~~~~~~~~~~~~~~~~~~~~$\rho_t \longrightarrow \infty,~~\mbox{as},~~z \longrightarrow \infty$,\\

~~~~~~~~~~~~~~~~~~~~~~~~~~~~~~$\rho_t \longrightarrow \rho^0+ \rho^1,~~\mbox{as},~~z \longrightarrow 0$,\\

~~~~~~~~~~~~~~~~~~~~~~~~~~~~~~$\rho_t \longrightarrow \rho^1,~~\mbox{as},~~z \longrightarrow -1$.\\
%%%%%%%%%%%%%%%%%%%%%%%%%%%%%%%%%%%%%%%%%%%%%%%%%%%%%%%%%%%%%%%%%%%%%%%%%%%%%%%%%%%%%%%%%%%%%%

As before, from the Eq. (\ref{variable3}), we see that, the present 
interacting DE (with variable equation of state) and DM is equivalent to a non-interacting
two fluid system with constant equations of state `0' (i.e., dust) and `$-n$' respectively.
The integration constants $\rho^0$ and $\rho^1$ are nothing but the energy
densities (at the present epoch) of the equivalent two fluid components.
Also, under the condition (\ref{variable2}), the solution for $\omega_d$ looks

\begin{equation}
\omega_d= \frac{n+ \lambda_d}{\omega_{d0} \left(1+ z\right)^{3 (n+ \lambda_d)}-1}, \label{variable4}
\end{equation}
%%%%%%%%%%%%%%%%%%%%%%%%%%%%%%%%%%%%%%%%%%%%%%%%%%%%%%%%%%%%%%%%%%%%%%%%%%%%%%%%%%%%%%%%%%%%
where $\omega_{d0}$ is the constant of integration, and it has been taken to be negative
to remove the singularity in $\omega_d$. Further, the graphical representations of
$\omega_d$ for different values of the interaction parameter ($\lambda_d$) have been presented
in FIG. 1, whereas FIG. 2 shows that in the high redshift era, $\omega_d$ was negative,
but very close to zero, thus, not dominating in nature, but it tracks $\rho_m$ after a
certain redshift to start a dark energy era. Further, at present, it is very close to the
$\Lambda$CDM ($\omega_d= -1$) for very small interaction parameter between the dark sectors,
and, also it matches with the very latest Planck data \cite{Planck-collaboration2015}.\\

\begin{figure}[h]
\begin{minipage}{0.45\textwidth}
\includegraphics[width= 0.70\linewidth]{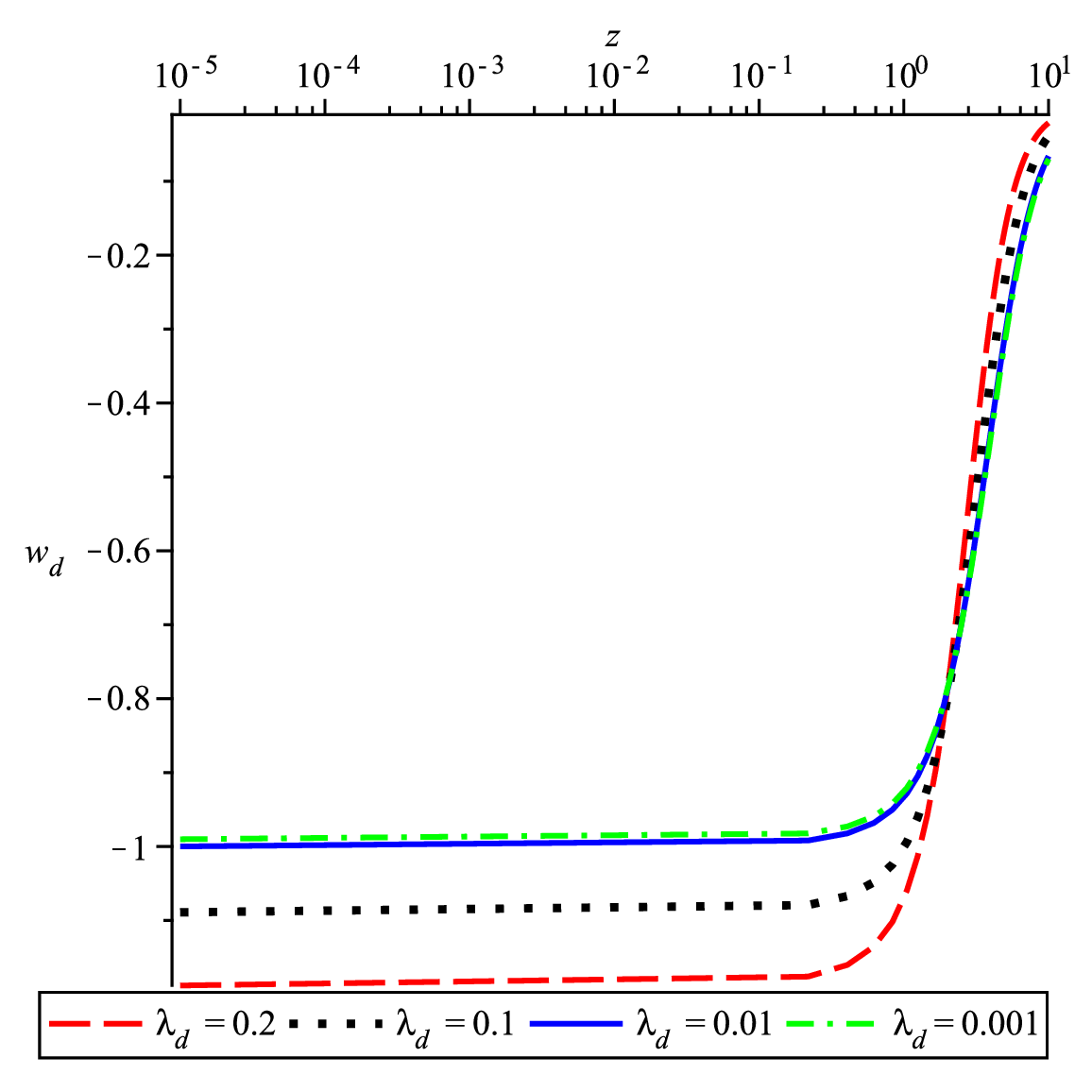}
\caption{The figure shows the behavior of the variable\\ $\omega_d$ throughout the entire evolution of the universe.}
\end{minipage}
\begin{minipage}{0.45\textwidth}
\includegraphics[width= 0.70\linewidth]{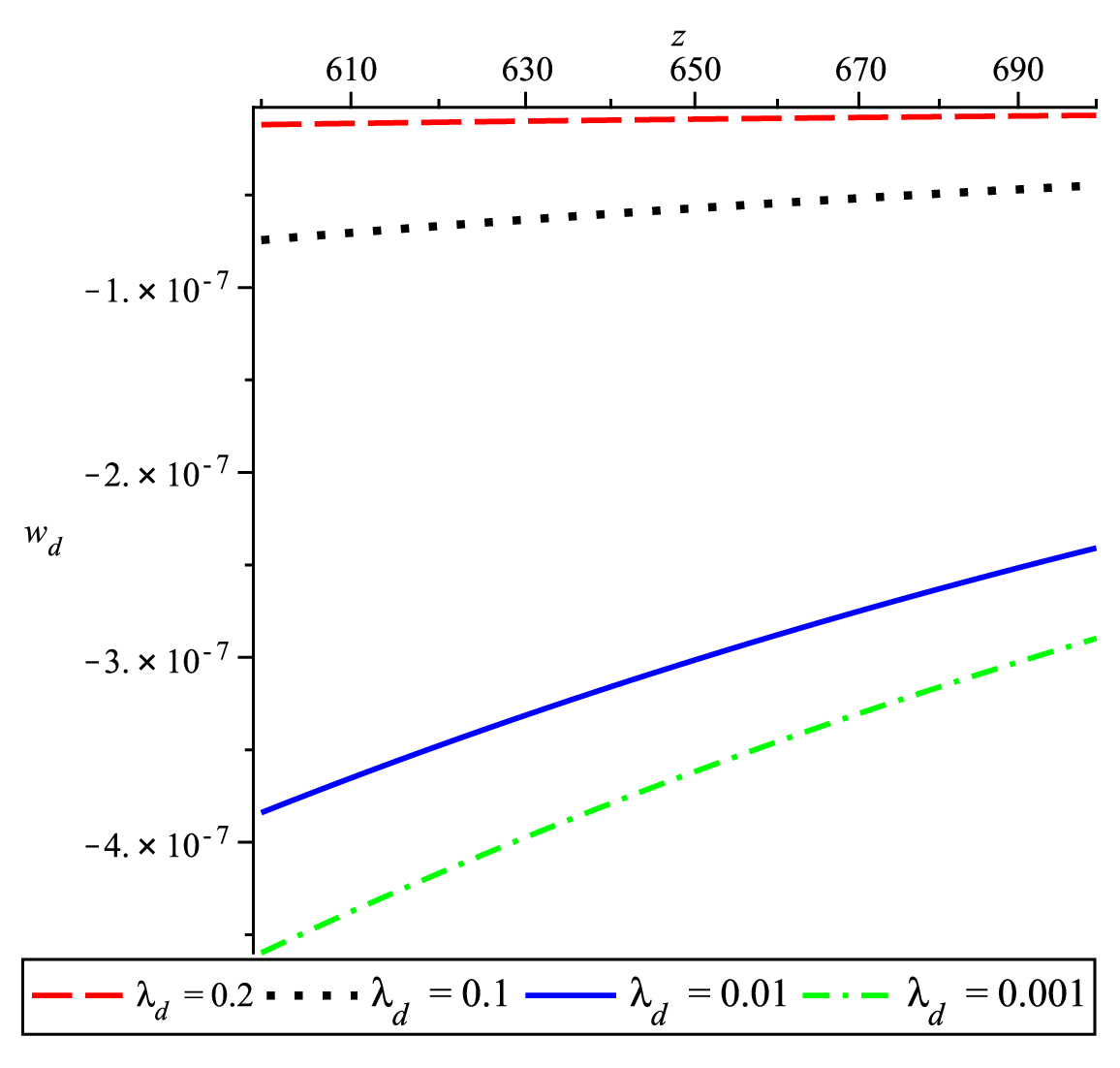}
\caption{This shows that in high redshift era, $\omega_d$ was still negative.}
\end{minipage}
\end{figure}

Moreover, we can give the explicit solutions for $\rho_m$ and $\rho_d$ as follows:

\begin{eqnarray}
\rho_m&=&\rho^0 (1+ z)^3+ \left(\frac{\rho^1 \lambda_d}{n+ \lambda_d}\right) (1+ z)^{3 (1-n)}+ \left(\frac{n \rho^1 \omega_{d0}}{n+ \lambda_d}\right) (1+ z)^{3 (1+ \lambda_d)},\label{variable-rhom}\\
\rho_d&=&\left(\frac{n \rho^1}{n+ \lambda_d}\right) \left[(1+ z)^{3 (1- n)}-\omega_{d0} (1+ z)^{3 (1+ \lambda_d)}\right].\label{variable-rhod}
\end{eqnarray}
%%%%%%%%%%%%%%%%%%%%%%%%%%%%%%%%%%%%%%%%%%%%%%%%%%%%%%%%%%%%%%%%%%%%%%%%%%%%%%%%%%%%%%
Similarly, for variable DE equation of state, the present day density parameters for DM 
($\Omega_{m0}$) and DE ($\Omega_{d0}$) can respectively be expressed as

\begin{eqnarray}
\Omega_{m0}&=& \Omega^ 0+ \left(\frac{\lambda_d+ n \omega_{d0}}{n+ \lambda_d}\right) \Omega^1,\label{variable-new1}\\
\Omega_{d0}&=& \left(1- \omega_{d0}\right) \left(\frac{n \Omega^1}{n+ \lambda_d}\right),\label{variable-new2}
\end{eqnarray}
%%%%%%%%%%%%%%%%%%%%%%%%%%%%%%%%%%%%%%%%%%%%%%%%%%%%%%%%%%%%%
where $\Omega^0= \rho^0/3 H_0 ^2$, $\Omega^1= \rho^1/3 H_0 ^2$. 
Also, the explicit forms of $\Omega^0$ and $\Omega^1$ are

\begin{eqnarray}
\Omega^0&=& \left(\frac{n+ \lambda_d}{n (1- \omega_{d0})}\right) \Omega_{d0},\label{variable-new3}\\
\Omega^1&=& \Omega_{m0}- \left(\frac{n \omega_{d0}+ \lambda_d}{n (1- \omega_{d0})}\right) \Omega_{d0},\label{variable-new4}
\end{eqnarray}
%%%%%%%%%%%%%%%%%%%%%%%%%%%%%%%%%%%%%%%%%%%%%%%%%%%%%%%%%%%%%%%%%%%
which immediately shows that, $\Omega_{m0}+ \Omega_{d0}= \Omega^0+ \Omega^1= 1$.\\

Now, we can give a comparative behavior of the energy densities for both dark energy
and dark matter in different eras throughout the entire evolution of the universe
as follows:\\

$\star$ {\bf $n> 1$:}\\

~~~~~~~~~~~~~~~~~~~~$\mbox{As}~z \longrightarrow \infty:~~\rho_d \longrightarrow \infty,~~\rho_m~~\mbox{is undefined}$,\\

~~~~~~~~~~~~~~~~~~~~$\mbox{As}~z \longrightarrow 0:~~\rho_d \longrightarrow \frac{n \rho^1}{n+ \lambda_d} (1- \omega_{d0}),~~\rho_m \longrightarrow \rho^1+ \frac{\rho^1 \lambda_d}{n+ \lambda_d}+ \frac{n \rho^1 \omega_{d0}}{n+ \lambda_d}$,\\

~~~~~~~~~~~~~~~~~~~~$\mbox{As}~z \longrightarrow -1:~~\rho_d \longrightarrow \infty,~~\rho_m \longrightarrow \infty$.\\

$\star$ {\bf $n< 1$:}\\

~~~~~~~~~~~~~~~~~~~~$\mbox{As}~z \longrightarrow \infty:~~\rho_d \longrightarrow \infty,~~\rho_m~~\mbox{is undefined}$,\\

~~~~~~~~~~~~~~~~~~~~$\mbox{As}~z \longrightarrow 0:~~\rho_d \longrightarrow \frac{n \rho^1}{n+ \lambda_d} (1- \omega_{d0}),~~\rho_m \longrightarrow \rho^1+ \frac{\rho^1 \lambda_d}{n+ \lambda_d}+ \frac{n \rho^1 \omega_{d0}}{n+ \lambda_d}$,\\

~~~~~~~~~~~~~~~~~~~~$\mbox{As}~z \longrightarrow -1:~~\rho_d \longrightarrow 0,~~\rho_m \longrightarrow 0$.\\

$\star$ {\bf $n= 1$:}\\

~~~~~~~~~~~~~~~~~~~~$\mbox{As}~z \longrightarrow \infty:~~\rho_d \longrightarrow \infty,~~\rho_m~~\mbox{is undefined}$,\\

~~~~~~~~~~~~~~~~~~~~$\mbox{As}~z \longrightarrow 0:~~\rho_d \longrightarrow \frac{n \rho^1}{n+ \lambda_d} (1- \omega_{d0}),~~\rho_m \longrightarrow \rho^1+ \frac{\rho^1 \lambda_d}{n+ \lambda_d}+ \frac{n \rho^1 \omega_{d0}}{n+ \lambda_d}$,\\

~~~~~~~~~~~~~~~~~~~~$\mbox{As}~z \longrightarrow -1:~~\rho_d \longrightarrow \frac{n \rho^1}{n+ \lambda_d},~~\rho_m \longrightarrow \frac{\lambda_d \rho^1}{n+ \lambda_d}$.\\

Further, the deceleration parameter can be given as \cite{Pan2013}

\begin{eqnarray}
q&=& \frac{1}{2}+ \frac{3}{2} \Omega_d \omega_d,\label{deceleration-variable}
\end{eqnarray}
%%%%%%%%%%%%%%%%%%%%%%%%%%%%%%%%%%%%%%%%%%%%%%%%%%%%%%%%%%%%%%%%%%%%%%%%%%%%%%%%%%%
where $\Omega_d$ can be found from Eq. (\ref{variable-rhod}). Now, looking at Eq. (\ref{variable4}) we see\\

{\bf A:} $\mbox{For}~n+ \lambda_d> 0$

\begin{eqnarray}
\mbox{For}~z&\longrightarrow& \infty,~~\omega_d~\longrightarrow~0~,\label{variable4.1}\\
\mbox{For}~z&\longrightarrow& 0,~~\omega_d~\longrightarrow~\left(\frac{n+ \lambda_d}{\omega_{d0} -1}\right),\label{variable4.2}\\
\mbox{For}~z &\longrightarrow& -1,~~\omega_d~\longrightarrow~ -(n+ \lambda_d).\label{variable4.2a}
\end{eqnarray}

{\bf B:} $\mbox{For}~n+ \lambda_d< 0$

\begin{eqnarray}
\omega_d &\longrightarrow& -(n+ \lambda_d),~\mbox{as}~z\longrightarrow \infty,\label{variable4.3}\\
\omega_d &\longrightarrow& \left(\frac{n+ \lambda_d}{\omega_{d0} -1}\right),~\mbox{as}~z\longrightarrow 0.\label{variable4.4}\\
\omega_d &\longrightarrow& 0,~~\mbox{as}~z\longrightarrow -1.\label{variable4.4a}
\end{eqnarray}

{\bf C:} $\mbox{For}~n+ \lambda_d= 0$\\

\begin{eqnarray}
\omega_d&=&- \frac{1}{3 ln \left(a/a_0\right)},~~a_0= \mbox{constant of integration}.\label{variable4.5}
\end{eqnarray}

\begin{eqnarray}
\mbox{Also,}~\omega_d \longrightarrow \frac{1}{3 ln a_0},~\mbox{as}~z \longrightarrow~0.
\end{eqnarray}
%%%%%%%%%%%%%%%%%%%%%%%%%%%%%%%%%%%%%%%%%%%%%%%%%%%%%%%%%%%%%%%%%
The cases A and B and C result the following table describing the
different phases of the universe restricted
by the model parameters.\\

$~~~~${\bf Table I:} The table shows the different Cosmic phases depending on the parameters.\\\\
\begin{tabular}{|c|c|c|c|}
\hline Types & Quintessence & $\Lambda$CDM & Phantom\\
\hline A~~~~& $1-3 (n+ \lambda_d)< \omega_{d0}< 1- (n+ \lambda_d)$~~~~~~~& $\omega_{d0}= 1- (n+ \lambda_d)$~~~~~~&~~~~~~$\omega_{d0}> 1- (n+ \lambda_d)$\\
\hline B~~~~& $1-3(n+ \lambda_d)< \omega_{d0} < 1- (n+ \lambda_d)$~~~~~~~& $\omega_{d0}=1- (n+ \lambda_d)$~~~~~~&~~~~~~$\omega_{d0}> 1- (n+ \lambda_d)$\\
\hline C~~~~& $1/e< a_0< \exp(-1/3)$~~~~~~~~& $a_0= \exp(-1/3)$~~~~~~~&~~~~$a_0\longrightarrow 1$\\
\hline
\end{tabular}\\\\

\begin{figure}[h]
%\begin{minipage}{0.45\textwidth}
\includegraphics[width= 0.5\linewidth]{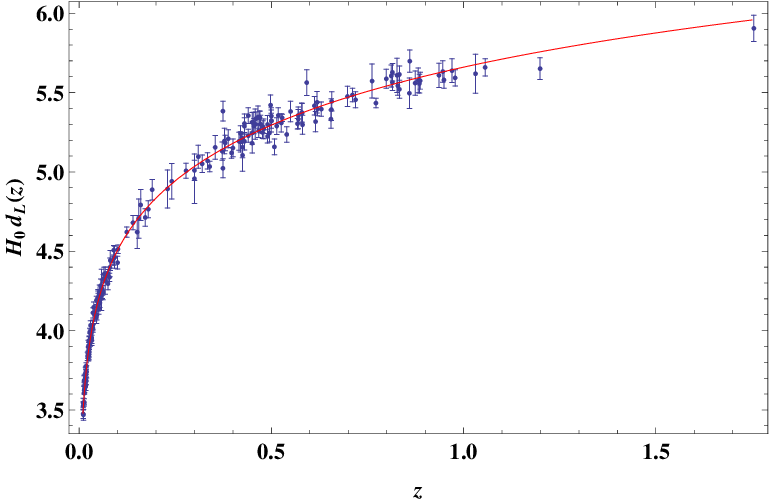}
\caption{The observed 194 Hubble free luminosity distance and the theoretically predicted luminosity distance (continuous graph) for $\omega$CDM (where $\omega= -1.01$) have been shown.}
%\end{minipage}
\end{figure}

\section{Model comparison with observational data}

Here we compare the models for both constant and variable $\omega_d$ up to the
redshift $z= 1.75$ using the available 194 Supernovae Ia data \cite{Tonry2003, Barris2004}.
The data is a compilation of the red shift $z$ and the corresponding logarithm
of the Hubble free luminosity distance $log (c~D_{L}(z))$ with its  $1\sigma$
error $\sigma_{log (D_{L}(z))}$. The Hubble constant free luminosity
distance $D_{L}(z)$ is related to the luminosity distance $d_{L}(z)$ by the relation

\begin{equation}
D_{L}(z)=\frac{H_{0}}{c}d_{L}(z).\label{observational1}
\end{equation}
%%%%%%%%%%%%%%%%%%%%%%%%%%%%%%%%%%%%%%%%%%%%%%%%%%%%%%%%%%%%%%%%%%%%%
In terms of the co-moving distance $r(z)$ and the red shift $z$,
$D_{L}(z)=\left(H_0/c\right) r(z) (1+ z)$. Again $D_{L}(z)$ can be
related to the theoretical model obtained using the relation

\begin{equation}
D_{L}^{th} (z)=\frac{(1+z)}{H_{0}}\int_{0}^{z}\frac{dz^\prime}{E(z^\prime)},\label{observational2}
\end{equation}
%%%%%%%%%%%%%%%%%%%%%%%%%%%%%%%%%%%%%%%%%%%%%%%%%%%%%%%%%%%%%%%%%%%%%%%%
where $E(z)=H(z)/H_0$. In order to determine the model parameters using
the observational constraints, we use a maximum likelihood technique on the
theoretical parameters whereby we minimize the function $\chi^{2}$ given by

\begin{equation}
\chi^{2}=\sum_{n=1}^{N}\frac{\left[log_{10}D_{L}^{obs}(z_{n})-log_{10}D_{L}^{th}(z_{n})\right]^{2}}{\left(\sigma_{log_{10}D_{L}^{obs}(z_{n})}\right)^{2}+ \left(\frac{\partial{log_{10}D_{L}^{obs}(z_{n})}}{\partial{z_n}}\sigma_{z_{n}}\right)^{2}},\label{observational3}
\end{equation}
%%%%%%%%%%%%%%%%%%%%%%%%%%%%%%%%%%%%%%%%%%%%%%%%%%%%%%%%%%%%%%%
where $N= 194$ and $\sigma_{z_{n}}$ is the $1 \sigma$ error of the data corresponding to the red shift $z_{n}$. A
table of the data and the numerical program we used in this study can be downloaded in electronic form
\cite{LP}

\begin{figure}[h]
\begin{minipage}{0.45\textwidth}
\includegraphics[width= 0.70\linewidth]{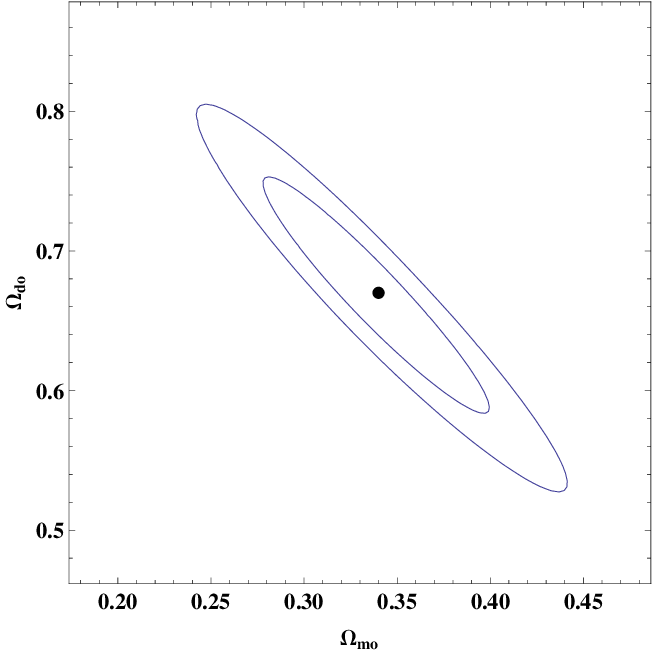}\\
\end{minipage}
\begin{minipage}{0.45\textwidth}
\includegraphics[width= 0.70\linewidth]{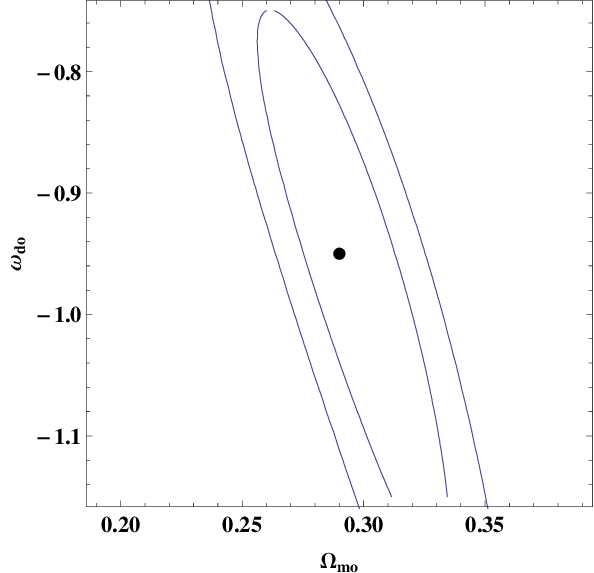}\\
\end{minipage}
\begin{minipage}{0.45\textwidth}
\includegraphics[width= 0.70\linewidth]{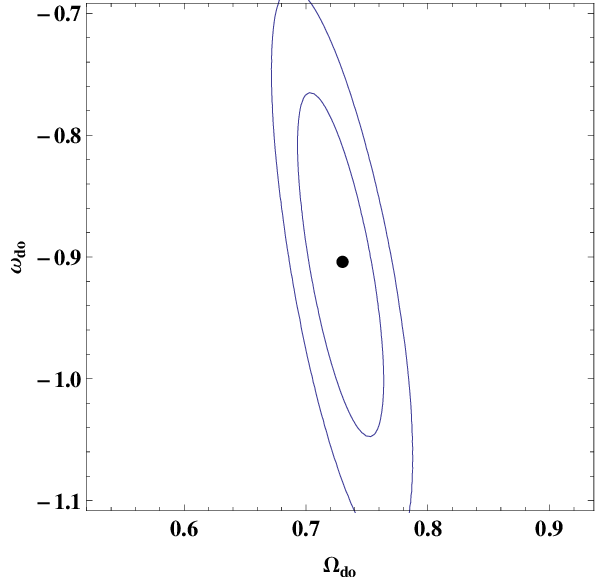}\\
\end{minipage}
\caption{68\% ($1\sigma$) and 95\% ($2\sigma$) confidence level contours in ($\Omega_{m0}, \Omega_{d0}$), ($\Omega_{m0}, \omega_{d0}$) and ($\Omega_{d0}, \omega_{d0}$) plane have been shown with the best fit parameter indicated by the black dot in each plot.}
\end{figure}

\subsection{Observational constraints for constant $\omega_d$}

From the Eq. (\ref{total-approx}), we can write

\begin{equation}
E^{2}=\Omega_{m0}(1+ z)^{3(1-\lambda_{m})}+\Omega_{d0} (1+ z)^{3(1+\omega_{d}+\lambda_{d})}.\label{E1}
\end{equation}
%%%%%%%%%%%%%%%%%%%%%%%%%%%%%%%%%%%%%%%%%%%%%%%%%%%%%%%%%%%%%%%%%%%%%%%%%%%%%%%%%%%%%%%
The parameters $\Omega_0$ and $\Omega_1$ are related with $\Omega_{m0}$ and $\Omega_{d0}$
given in (\ref{neweqn3}) and (\ref{neweqn4}) respectively. Further, $\Omega_0$ and $\Omega_1$
can be interpreted as the density parameters of the equivalent two fluids with 
corresponding values equivalent to that of $\Omega_{m0}$ and $\Omega_{d0}$ respectively.\\

FIG. 3 is a representation of the observed luminosity distance for the 194 Sne Ia
data and the corresponding model predicted theoretical value for $d_{L}(z)$. For
evaluating theoretical $d_{L}(z)$ we use $\omega$CDM model parameters for
$\Omega_{m0}= 0.34$ and $\omega_{d}= -1.01$, i.e., we consider a small deviation
of the $\Lambda$CDM model, and, thus we get $\Omega_{d0}= 0.66$. We choose the
interaction parameters, $\lambda_{m}= 0.001$ and $\lambda_{d}= 0.002$ throughout
all estimations as per our assumption of very low interaction. From the figure it
is clear that for these values of the parameter, our model gives a good fit to the
data.

FIG. 4 shows the $1\sigma$ and $2\sigma$ contours in $\Omega_{m0}, \Omega_{d0}$
plane, $\Omega_{m0}, \omega_{d0}$ and $\Omega_{d0}, \omega_{d0}$ plane respectively.
In all these results we have found that the best fit values of the free parameters
$\Omega_{m0}/\Omega_{d0}$ and $\omega_{d}$ are consistent with the data. Thus,
from the observational constraints we can conclude that the long term expansion
history of the universe is overall in harmony with the existing models like $\Lambda$CDM.

\begin{figure}[h]
\begin{minipage}{0.45\textwidth}
\includegraphics[width= 0.70\linewidth]{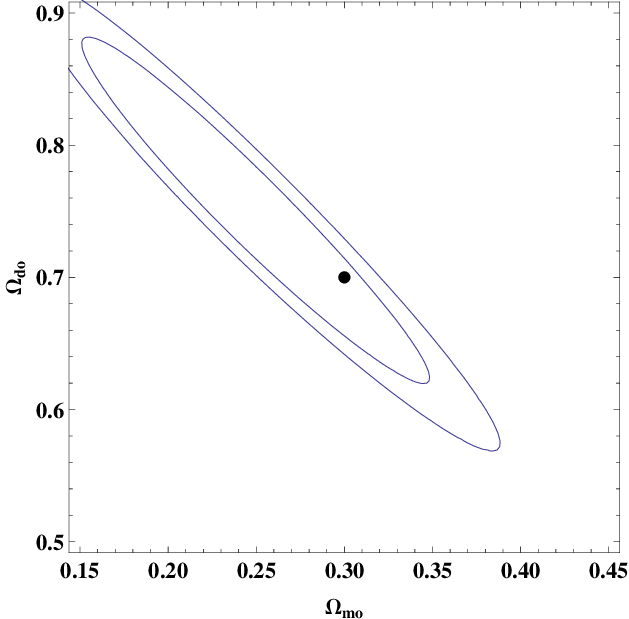}
\end{minipage}
\begin{minipage}{0.45\textwidth}
\includegraphics[width= 0.70\linewidth]{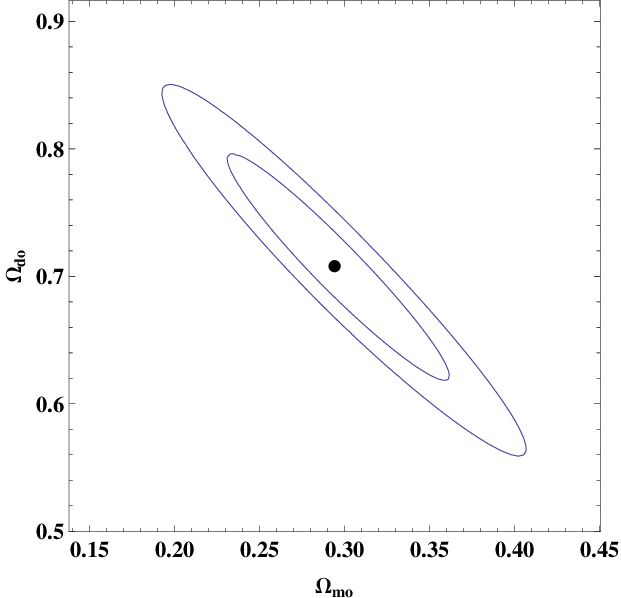}
\end{minipage}
\begin{minipage}{0.45\textwidth}
\includegraphics[width= 0.70\linewidth]{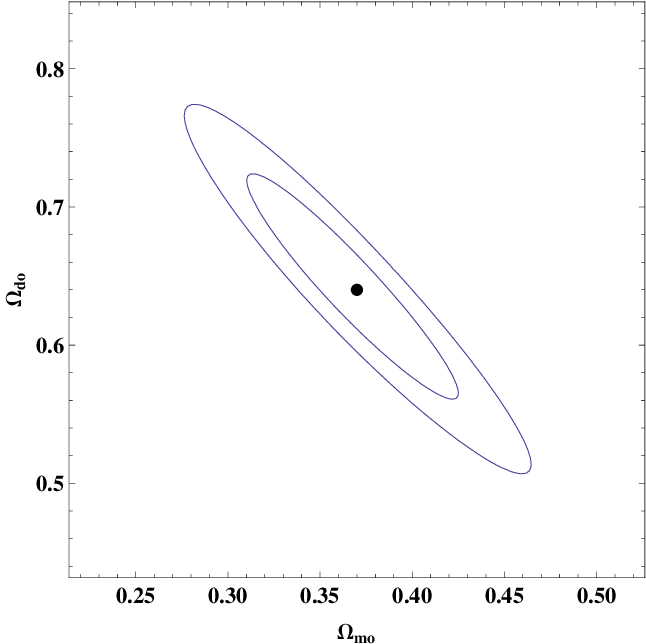}
\end{minipage}
\caption{The $1\sigma$ and $2\sigma$ confidence levels in the
($\Omega_{m0}, \Omega_{d0}$) plane have been shown for $n= 0.8$,
$n= 0.9$, and $n= 1.1$ respectively in the clockwise direction.
The best fit values of ($\Omega_{m0}, \Omega_{d0}$) for different
$n$ have also been indicated by the black dot in each figure.}
\end{figure}

\subsection{Observational constraints for variable $\omega_d$}

For variable $\omega_d$ restricted by Eq. (\ref{variable2}), we have

\begin{equation}
E^{2}=\Omega_{m0}(1+ z)^{3}+\Omega_{d0}(1+ z)^{3(1-n)}.\label{E2}
\end{equation}
%%%%%%%%%%%%%%%%%%%%%%%%%%%%%%%%%%%%%%%%%%%%%%%%%%%%%%%%%%%%%%%%%%%%%%%%
where $\Omega^0$ and $\Omega^1$ can be found in (\ref{variable-new3}) and 
(\ref{variable-new4}) respectively. In this case also, the parameters $\Omega^0$ and 
$\Omega^1$ are nothing but the density parameters for the hypothetical non-interacting two fluids
with values similar to $\Omega_{m0}$ and $\Omega_{d0}$ respectively. 
The new parameter $n$ (any real number) arises due to the choice of $\omega_{d}$
that makes it variable and as a result we can realize the different cosmic stages
with the restrictions on the model parameters shown in Table I. FIG. 5 shows
the $1\sigma$ and $2\sigma$ contours in $\Omega_{m0}, \Omega_{d0}$ plane for
three different values of $n$.

\section{Cosmography of interacting dark energy}

The idea of cosmography in cosmology was motivated after the
introduction of the statefinder parameters by Sahni et al. \cite{Sahni2003}.  The
interesting fact behind the statefinder parameters are that, they are dimensionless
geometrical, and model independent in nature. As a result, they were
widely used to filter the observationally sound dark energy
models among the various theoretical DE models in the literature.
The statefinder parameters $\{r, s\}$ are defined as

\begin{eqnarray}
r&=& \frac{1}{aH^3} \frac{d^3a}{dt^3},~~~\mbox{and,}~~s=\frac{r-1}{3 \left(q- \frac{1}{2}\right)}.\label{cosmographic-new1}
\end{eqnarray}
%%%%%%%%%%%%%%%%%%%%%%%%%%%%%%%%%%%%%%%%%%%%%%%%%%%%%%%%%%%%%%%%%%%%%%%%%%%%%%%%%%
Subsequently, this geometric investigation was
extended by considering the Taylor series expansion of
the scale factor about the present time in the following manner:

\begin{figure}[h]
\begin{minipage}{0.45\textwidth}
\includegraphics[width= 0.75\linewidth]{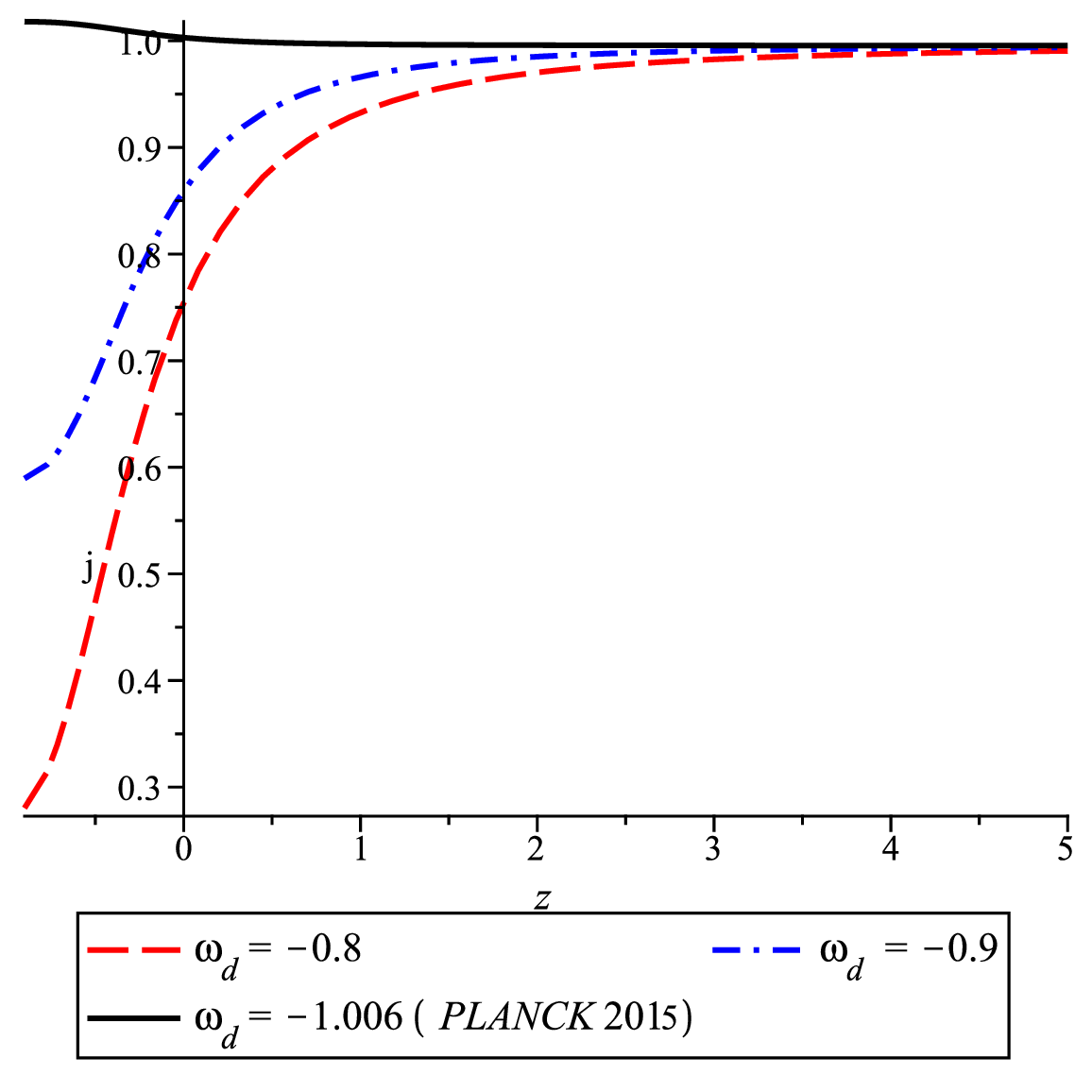}
\end{minipage}
\begin{minipage}{0.45\textwidth}
\includegraphics[width= 0.75\linewidth]{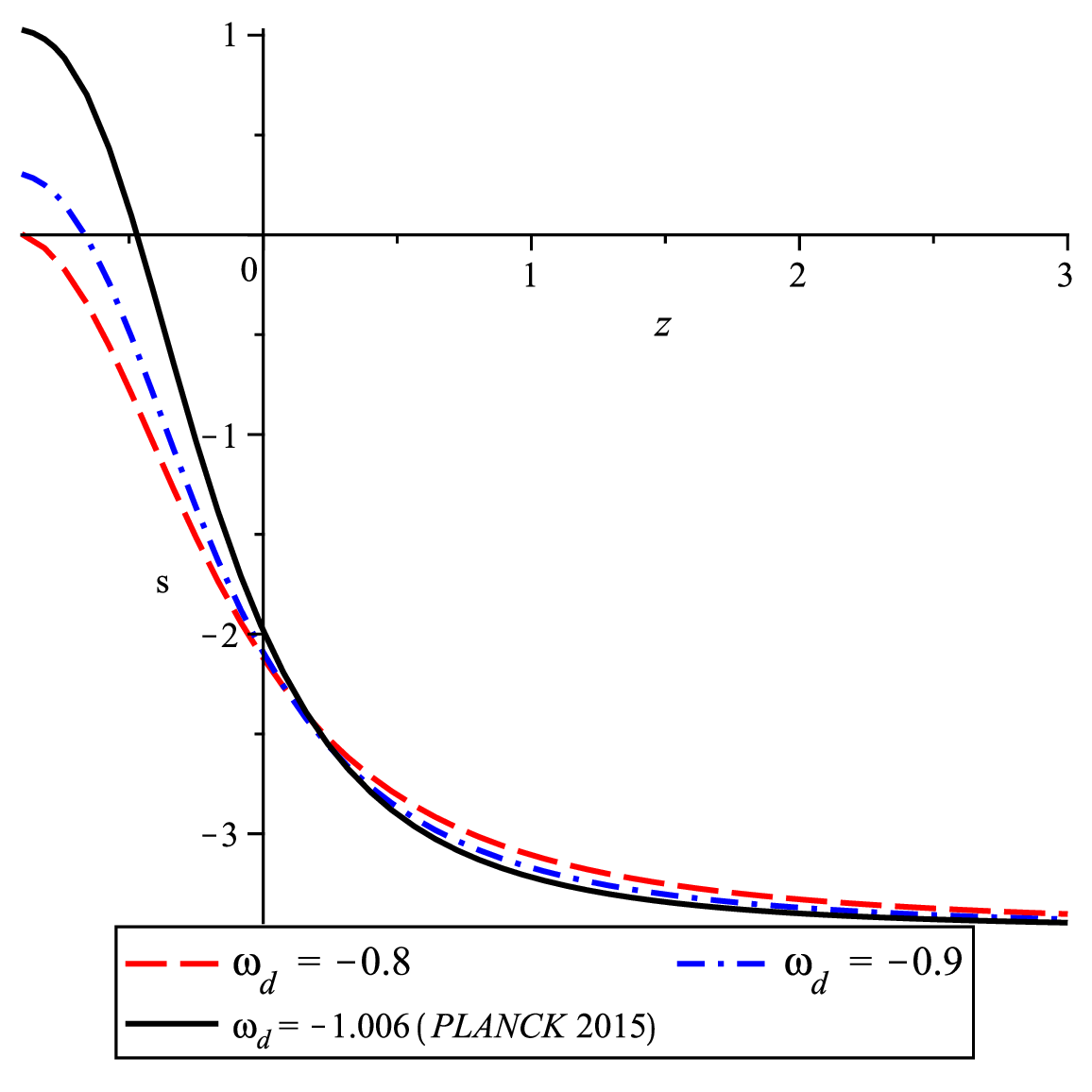}
\end{minipage}
\begin{minipage}{0.45\textwidth}
\includegraphics[width= 0.75\linewidth]{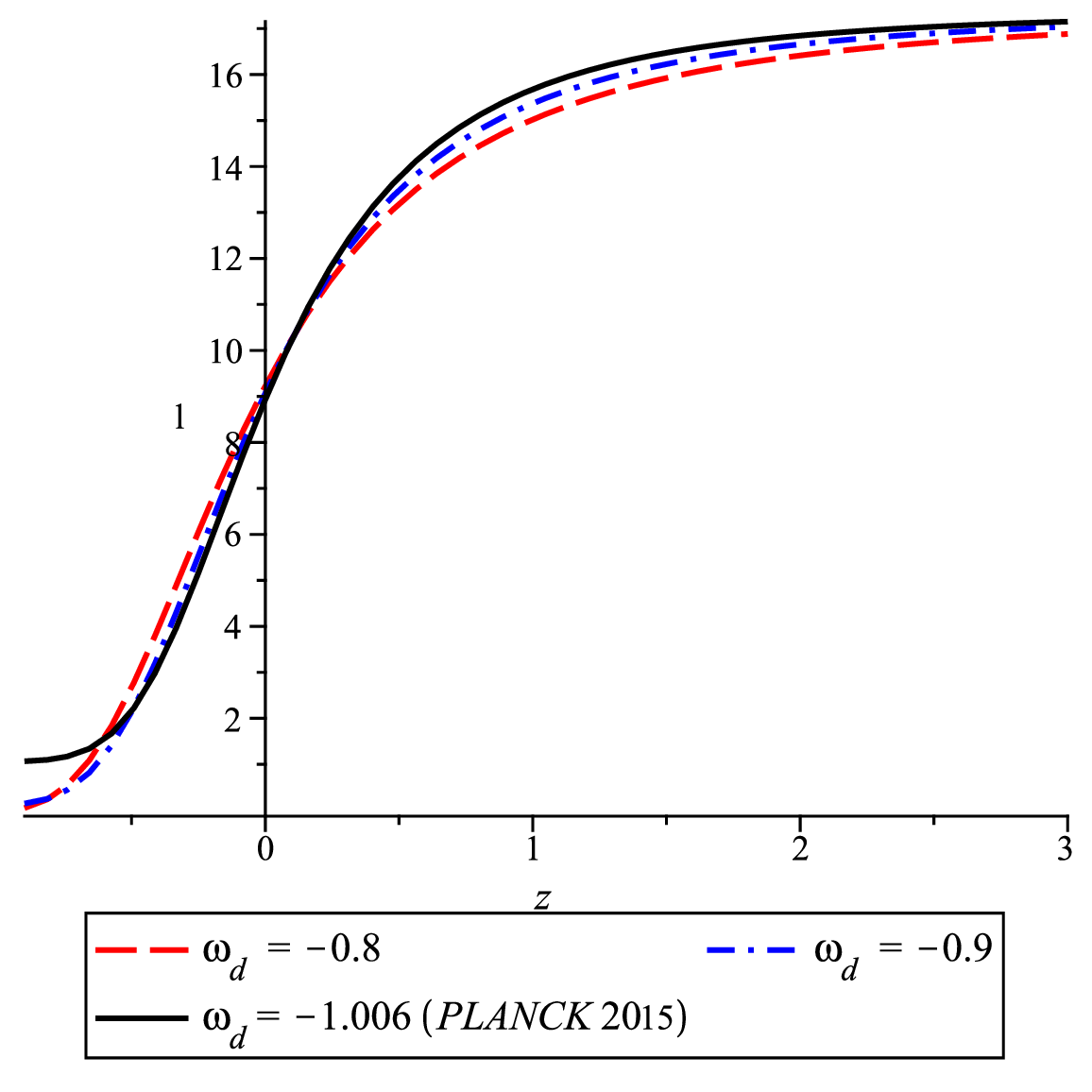}
\end{minipage}
\begin{minipage}{0.45\textwidth}
\includegraphics[width= 0.75\linewidth]{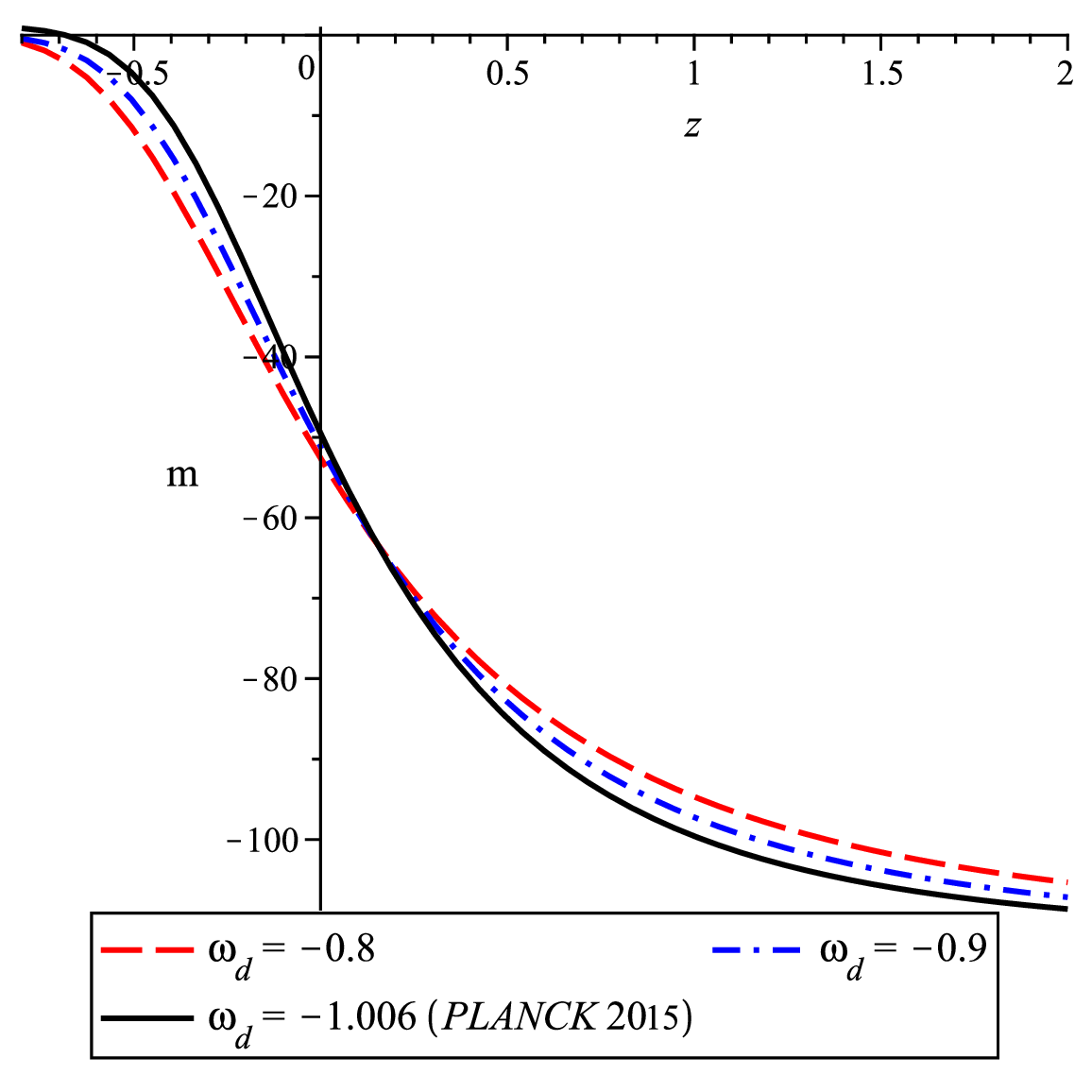}
\end{minipage}
\caption{The figures show the variation of the 4 CP against
the redshift ($z$) for constant EoS ($\omega_d$) for dark energy }
\end{figure}

\begin{eqnarray}
\frac{a (t)}{a(t_0)}&=& 1+ H_p (t-t_0)+ \frac{1}{2!} q_0 H_0 ^2 (t-t_0)^2+ \frac{1}{3!} j_0 H_0 ^3 (t-t_0)^3+ \frac{1}{4!} s_0 H_0 ^4 (t-t_0)^4+O[(t-t_0)^5],\label{sp-power-series}
\end{eqnarray}
%%%%%%%%%%%%%%%%%%%%%%%%%%%%%%%%%%%%%%%%%%%%%%%%%%%%%%%%%%%%%%%%%%%%%%%%%%%%%%%%5
where we have some model independent and dimensionless parameters $j, s, l, m$ known as the cosmographic
parameters \cite{Visser2004, Visser2005} defined in the following way:

\begin{eqnarray}
j&=& \frac{1}{aH^3} \frac{d^3 a}{dt^3},~~s= \frac{1}{aH^4} \frac{d^4 a}{dt^4},~~l= \frac{1}{aH^5}\frac{d^5 a}{dt^5},~~\mbox{and},~~m= \frac{1}{aH^6}\frac{d^6 a}{dt^6}.\label{sp-CP-parameters}
\end{eqnarray}
%%%%%%%%%%%%%%%%%%%%%%%%%%%%%%%%%%%%%%%%%%%%%%%%%%%%%%%%%%%%%%%%%%%%%%%%%%%%%%
The suffix `$0$' stands for the value of the corresponding variable at the
present epoch ($t_0$). The cosmographic parameters (from now we shall call these CP)
are individually named as jerk ($j$) (this `$j$' is same as `$r$'
defined by Sahni et al. \cite{Sahni2003}), snap ($s$) (this `$s$' is different from
one defined by Sahni et al. \cite{Sahni2003}), lerk, and $m$
parameter \cite{Visser2004, Visser2005}. Further, the above CP can be expressed in
terms of the deceleration parameter ($q$), and, its higher derivatives:

\begin{eqnarray}
j&=&  ~~(1+ z) \frac{dq}{dz}+ q (1+ 2 q),\label{form-j}\\
s&=& - (1+ z)\frac{dj}{dz}+ j- 3 (1+ q)j,\label{form-s}\\
l&=& - (1+ z)\frac{ds}{dz}+ s- 4 (1+q)s,\label{form-l}\\
m&=& - (1+ z)\frac{dl}{dz}+ l- 5(1+q)l\label{form-m}
\end{eqnarray}
%%%%%%%%%%%%%%%%%%%%%%%%%%%%%%%%%%%%%%%%%%%%%%%%%%%%%%%%%%%%%%%%%%%%%%%%%%%
Thus, the evolution of the CP can be traced out only if $q$ and its higher derivatives are differentiable
throughout the entire cosmic history. In FIGs. 6 and 7, we have shown the variations of the CP with the
evolution of universe for constant and variable equation of state parameters respectively. From the figures,
one can notice that during the entire evolution of the universe, the nature of $j$ and $l$ are almost same in sign,
similar behavior is found for the parameters $s$ and $m$.

\begin{figure}[h]
\begin{minipage}{0.45\textwidth}
\includegraphics[width= 0.85\linewidth]{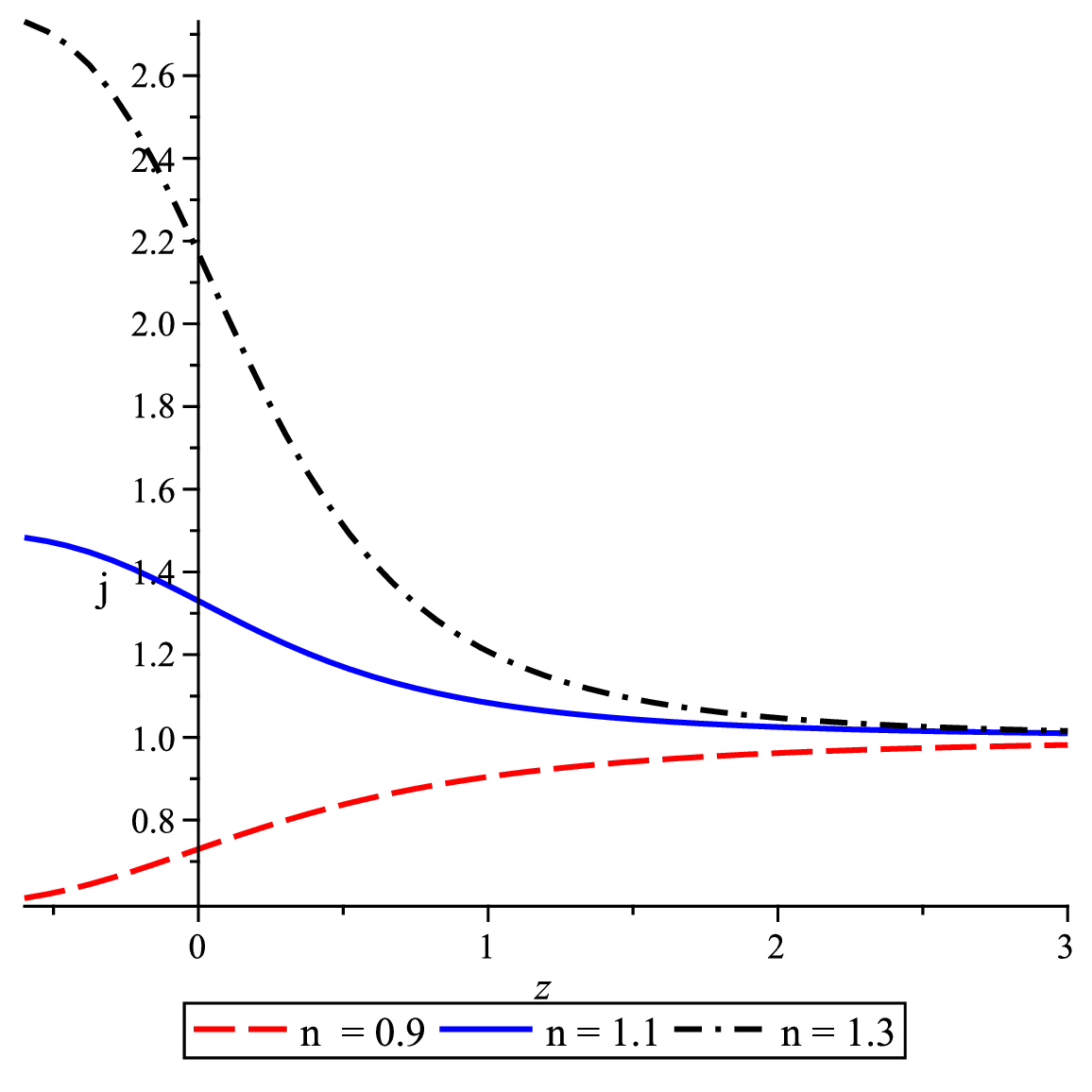}
\end{minipage}
\begin{minipage}{0.45\textwidth}
\includegraphics[width= 0.85\linewidth]{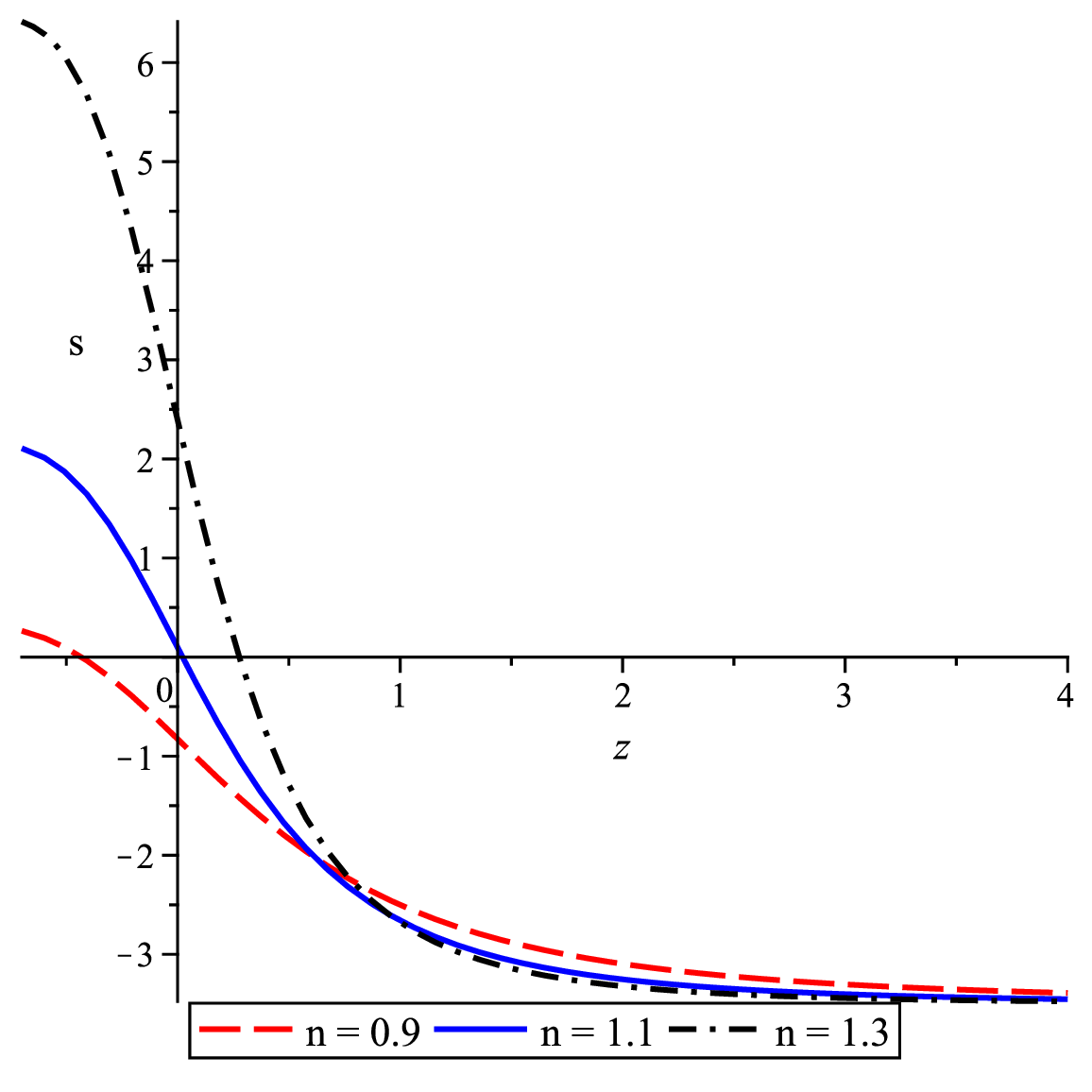}
\end{minipage}
\begin{minipage}{0.45\textwidth}
\includegraphics[width= 0.85\linewidth]{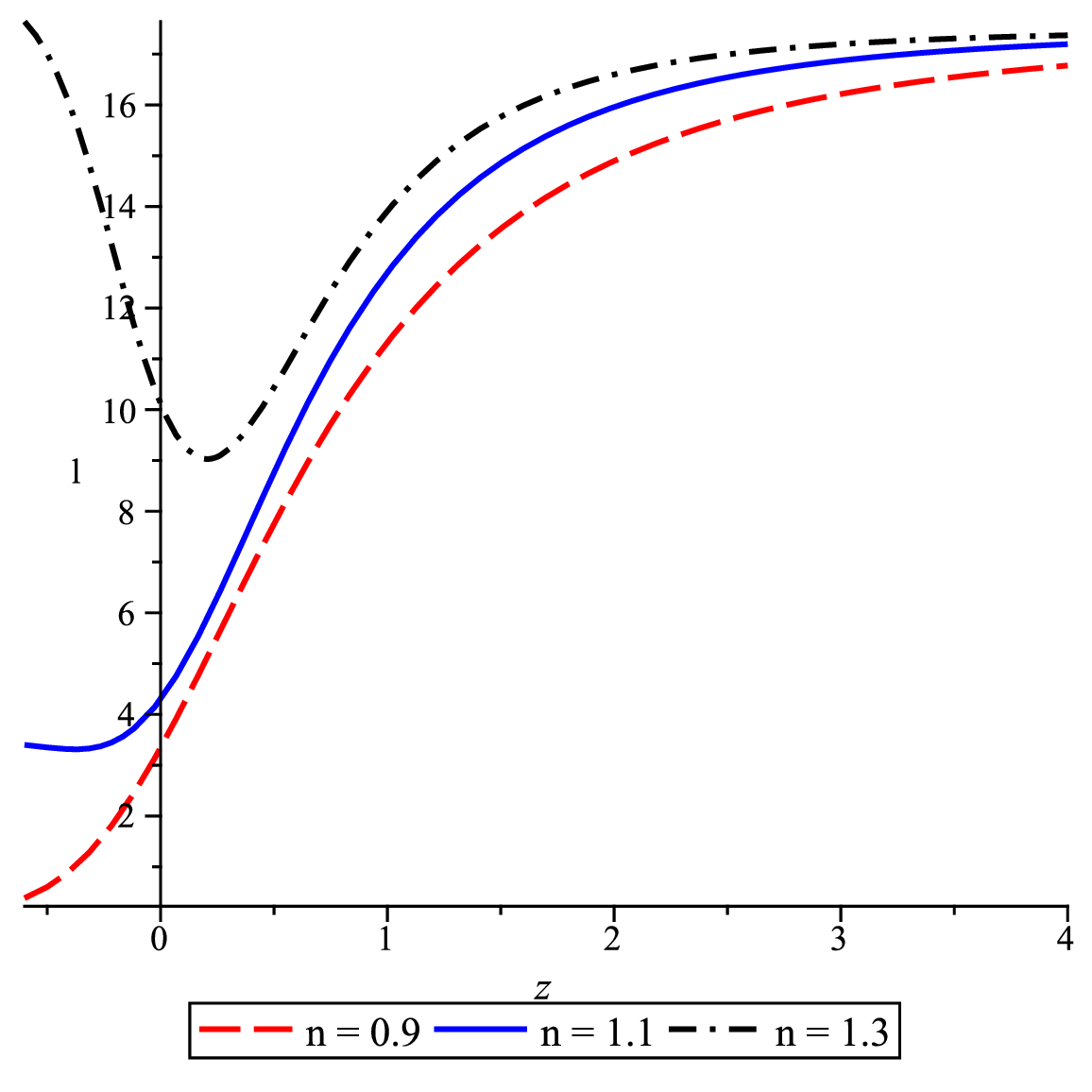}
\end{minipage}
\begin{minipage}{0.45\textwidth}
\includegraphics[width= 0.85\linewidth]{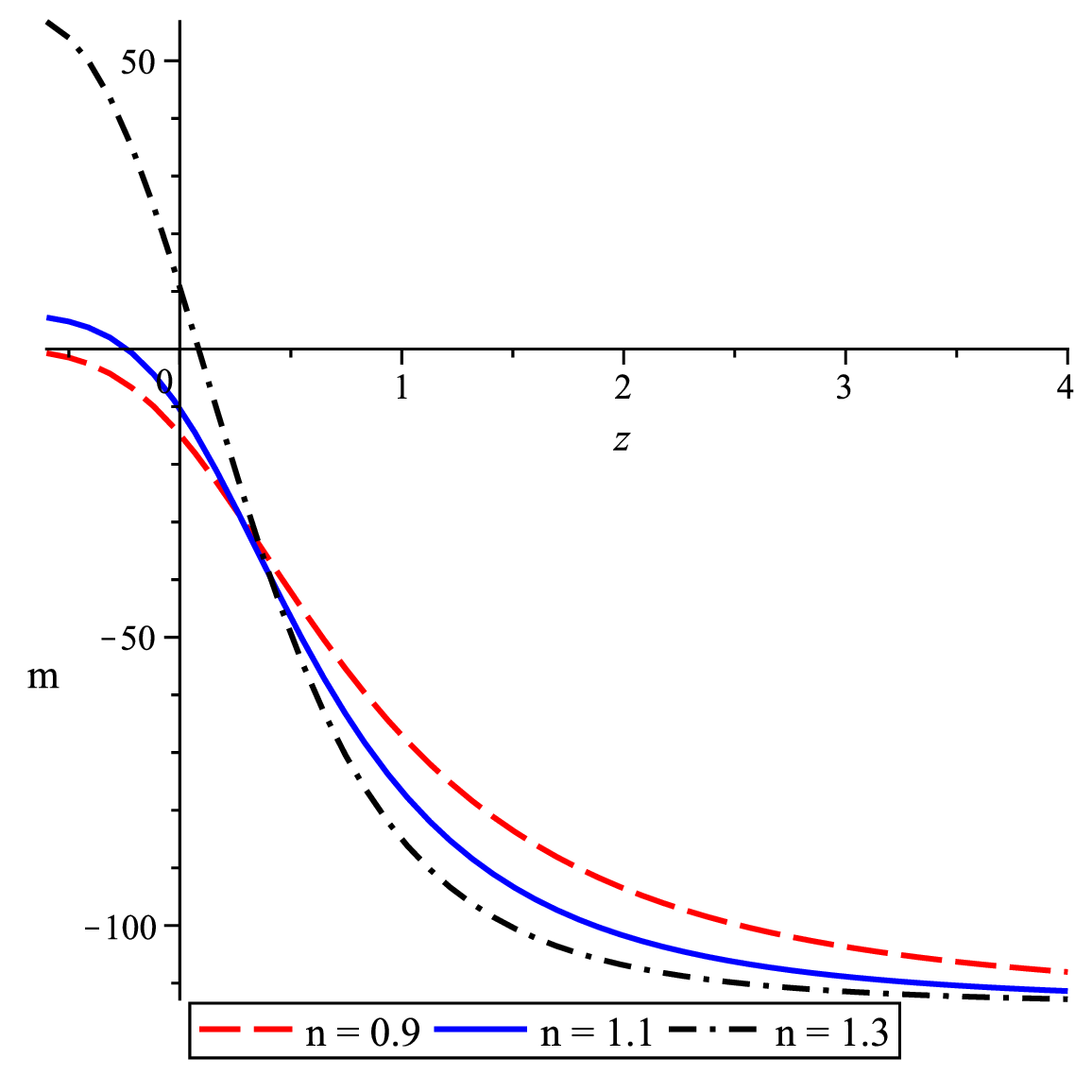}
\end{minipage}
\caption{For variable EoS of dark energy ($\omega_d$) in Eq.
(\ref{variable4}), we have presented the CP against the redshift ($z$).}
\end{figure}

\section{Summary of the work}

The present work proposes a theoretical model of dark energy to match
with recent observational evidences. Here, the dark energy is chosen
in the form of perfect fluid with barotropic equation of state, and, it
interacts non-gravitationally with dark matter chosen as dust. Analytic
solutions are obtained both for constant and variable equation of state
for DE with phenomenological choice for the interaction term. The asymptotic
behavior of the relevant physical parameters are discussed and their variations
have been shown graphically. These theoretical models are then compared with
194 available Supernovae data by Tonry et al. \cite{Tonry2003}
and Barris et al. \cite{Barris2004}. Using $\chi^2$ test for goodness of fit,
the best fit values for the parameters are estimated, and, they are well
accord with observed 1$\sigma$ (or, 2$\sigma$) level. For variable $\omega_d$,
the effect of DE is negligible at early universe, and, it starts dominating in
recent past. Also, the estimated value of $\omega_d$ nicely matches with the
very recently released Planck data set \cite{Planck-collaboration2015}. Therefore,
this particular interacting DE model can be considered as an alternative for
$\Lambda$CDM model. Finally, the variation of the cosmographic parameters are
presented graphically for both the theoretical models for different choices
of the parameters involved. Finally, as the
present model includes only CDM and DE as the two main ingredients,
one can include the baryonic matter and radiation to the picture
\cite{Erdem} in future such that the era of the creation of baryonic
matter and radiation is built into the model.

%We have discussed dark energy interacting with dust through a phenomenological
%interaction term considering both constant and variable EoS for dark energy.
%We have introduced an ansatz for variable $\omega_d$. Both the cases have been
%analyzed with 194 available Supernovae data by Tonry et al. \cite{Tonry2003}
%and Barris et al. \cite{Barris2003}. For constant EoS for dark energy, we have
%the best fit parameters: $(\Omega_{m0}, \Omega_{d0})= (0.34, 0.66)$ with $\chi^2= 198.616$;
%$(\Omega_{m0}, \omega_{d0})= (0.30, -0.95)$ with $\chi^2= 199.297$, and
%$(\Omega_{d0}, \omega_{d0})= (0.73, -0.90)$ with $\chi^2= 199.487$.
%In case of variable $\omega_d$, it is seen that, at early time,
%it was very very negligible to execute a dark energy universe, but later on,
%it tracks the matter energy density at some redshift, and, finally, at late-time,
%the model behaves exactly with the $\Lambda$CDM model, and also, it agrees
%with the latest released Planck data set \cite{Planck-collaboration2015},
%where $\omega_d= -1.006 \pm 0.045$. So, this particular interacting dark
%energy model can be considered as an alternative for $\Lambda$CDM model.
%We have also presented a cosmographic behavior of the different cosmographic
%parameters both for constant and variable equation of state for dark energy.

\section{Acknowledgments}
SP acknowledges CSIR, Govt of India for financial support 
through SRF scheme (File No. 09/096 (0749)/2012-
EMR-I). SB acknowledges UGC's Faculty Recharge Programme. 
SC thanks UGC-DRS programme at the Department of 
Mathematics, Jadvapur University. SP thanks R.
Erdem for some critical comments on the manuscript. SP
also thanks S. Das for useful discussions in understanding 
the numerical programs. Finally, the authors are
thankful to the referee whose critical comments helped
us to improve the manuscript considerably.

\end{document}